\documentstyle[12pt]{article}

\newcommand{\be}{\begin{equation}}
\newcommand{\ee}{\end{equation}}
\newcommand{\bea}{\begin{eqnarray}}
\newcommand{\eea}{\end{eqnarray}}

\newcommand{\g} {{\overline g}}
\newcommand{\p} {{\overline \phi}}
\newcommand{\V} {{\overline V}}
\newcommand{\G} {{\tilde g}}
\newcommand{\pp} {{\tilde \phi}}
\newcommand{\VV} {{\tilde V}}

\begin{document}

\begin{center}
\begin{large}
{\bf  Quantum-Corrected Cardy Entropy \\}
{\bf  for \\}
{\bf  Generic 1+1-Dimensional Gravity  \\}
\end{large}  
\end{center}
\vspace*{0.50cm}
\begin{center}
{\sl by\\}
\vspace*{1.00cm}
{\bf A.J.M. Medved\\}
\vspace*{1.00cm}
{\sl
Department of Physics and Theoretical Physics Institute\\
University of Alberta\\
Edmonton, Canada T6G-2J1\\
{[e-mail: amedved@phys.ualberta.ca]}}\\
\end{center}
\bigskip\noindent
\begin{center}
\begin{large}
{\bf
ABSTRACT
}
\end{large}
\end{center}
\vspace*{0.50cm}
\par
\noindent

Various studies have  explored the possibility of
explaining the  Bekenstein-Hawking (black hole) entropy
by way of  some suitable  state-counting procedure. Notably, many of
these treatments have used the well-known Cardy formula
as an intermediate step. Our current  interest is 
a recent calculation  in which Carlip has  deduced
the  leading-order quantum
 correction to the (otherwise) classical Cardy formula.
In this paper, we apply Carlip's  formulation
to the case of a generic model of two-dimensional
gravity with coupling to a dilaton field. We find
that the corrected Cardy entropy  includes the anticipated logarithmic
``area'' term. Such a term is also evident
when the entropic correction is  derived independently   by thermodynamic
means. However, there is an apparent
 discrepancy between the two calculations
with regard to the factor in front of the logarithm. 
In fact, the two values of this prefactor can only agree for very
specific
two-dimensional models, such as that describing Jackiw-Teitelboim theory.

\newpage

\par

\section{Introduction}

It is safe to say that the Bekenstein-Hawking definition
of black hole entropy \cite{bek,haw}\footnote{For future reference, 
the Bekenstein-Hawking  entropy
is defined as one quarter of the black hole horizon area
(or the analogue of area when the spacetime dimensionality differs from
four) 
 divided
by Newton's constant. Here and throughout, 
all other fundamental constants  have been set equal to unity.} 
has become a fixture of gravitational
physics.
 It is commonly believed that any valid  theory of quantum gravity
must necessarily incorporate this entropy
into its conceptual framework. What is, however, still absent (along
with the elusive theory of quantum gravity itself), is some
sort of statistical-mechanical explanation for this entropy,
which has its genesis in thermodynamic principles. That is
to say, the microscopic origin of black hole entropy (assuming
there are indeed microscopic degrees of freedom that underlie this quantity)
remains one of the most profound open questions in theoretical physics.  
\par 
In spite of the above statements, there have been  many ingenious  
attempts at calculating this entropy via statistical means;
a significant number of which have enjoyed dramatic 
success. 
These include the following partial list: calculations based
on string and D-brane theories \cite{str2},  quantum geometry \cite{ash},
a Chern-Simons formulation of 2+1-dimensional gravity \cite{carlip3},
Sakharov-inspired \cite{sak} induced gravity \cite{fro,fur2},
conformal symmetries at spatial infinity for
2+1-dimensional \cite{str} and 1+1-dimensional 
gravity \cite{cad3,cadx,cad}, and  conformal symmetries at the black hole 
horizon for  1+1-dimensional gravity \cite{fur}
and for  arbitrary dimensionality \cite{carlip5,sol,carlip}.
(For a general overview, see Ref.\cite{carlip2}.)
\par
The above list is indeed impressive and  provides a  strong indication  
that such research has been heading in the
right direction. However, the microscopic origin of entropy remains an enigma
for (at least) two reasons. First of all, 
although the various counting methods have pointed
to the expected semi-classical result, there is still a lack of
recognition as to what degrees of freedom are truly being counted.
This ambiguity  can be attributed to  most of these methods being based
on dualities with simpler theories; thus
 obscuring the physical interpretation from
the perspective of the black hole in question.
Secondly, the vast and varied number of successful counting techniques 
only serve to  cloud up an  already fuzzy  picture. One would hope
for some sort of underlying, universal principle to be at work, but it remains
a mystery as to what this may be.
\par 
As is often the case in resolving difficulties  in physics,
it is useful to reformulate the problem in terms of  a lower-dimensional 
theory.
This has, in large part, been the motivation
for the above-cited studies with regard to 2+1 and 1+1-dimensional
gravity. However, along with the desirable feature of simplicity, such models
often have physical significance via dual relationships
with higher-dimensional theories. For instance, consider 2+1-dimensional
anti-de Sitter gravity, which is known to admit  
BTZ black hole solutions \cite{btz}.
This theory has  turned  out to be
 relevant to many string-theoretical black holes,  for which 
the near-horizon geometries take on the form of BTZ times a simple
manifold \cite{youm}.  Another example is 1+1-dimensional anti-de Sitter
gravity,
which admits   black hole solutions that are described by Jackiw-Teitelboim
theory \cite{jt}.  This two-dimensional model  has also been 
shown to have dual relationships with  
certain string-inspired black holes \cite{youm}. Furthermore,  
the Jackiw-Teitelboim  model  can be used to  effectively describe
the near-horizon geometry of
higher-dimensional, near-extremal black holes (such as the near-extremal
Reissner-Nordstrom solution) \cite{ext}. 
\par
To further motivate the study of gravity in two dimensions of spacetime,
we  point out that
such theories can also arise from an
appropriate reduction of a higher-dimensional theory. This includes
the spherically symmetric reduction of Einstein gravity \cite{spher}
and a reduction of the BTZ model that leads to Jackiw-Teitelboim
theory \cite{ao}.
With regard to the specific problem of explaining the microscopic
origin of black hole entropy,  two-dimensional theories have
an added appeal on their own right. It has been argued that,
in the context of a   1+1-dimensional black hole, 
the degrees of freedom  being counted by the Cardy formula
(see below) may actually represent the physical microscopic states
of the underlying theory \cite{fur,fur2}. Conversely,  such a direct physical
interpretation  of the Cardy formula
  seems to be lacking in the case of higher-dimensional black
holes.
\par  
Let us, for the moment, put aside
 the topic of dimensionally reduced theories 
and return our focus to the procedure of microstate counting.
Many of the priorly cited studies  are essentially
 based on the following premise.
The geometry of the black hole theory in question
 effectively behaves as a two-dimensional
conformal field theory at a suitable boundary; either at spatial infinity
or near the horizon. This enables one, in principle,
 to evaluate the exponent of the  black hole entropy  by  counting 
the 
states of the dual boundary theory.  Such an evaluation is possible
via Cardy's well-known formula for the density of states of
a two-dimensional conformal field theory \cite{car}. For
an explicit calculation, two ingredients are needed: 
 the  ``central charge'' and the eigenvalue
of the  zero-mode  generator  of the corresponding
Virasoro algebra (which can be used to describe
the symmetries of a  conformal field theory \cite{cft}).
In principle, one can evaluate these quantities by
identifying the  symmetries at the boundary and then 
formulating the relevant generators so that they explicitly
realize the Virasoro commutator relations. 
\par
Of particular interest to the current paper is a (relatively) recent 
study by Carlip \cite{carlip}. This author was able to calculate 
the leading-order  quantum correction to the (otherwise) classical
Cardy expression. The revised formula was then tested for
several specific cases; for instance,  the entropy of a BTZ black hole 
as based on a  Virasoro algebra that was identified by   
Strominger \cite{str}
(also see Ref.\cite{brown}). In all of these cases, the correction was
found  (up to an irrelevant constant) to be  proportional to the logarithm of
the horizon area (or, equivalently, the logarithm of
the Bekenstein-Hawking entropy). Moreover, the prefactor
of this logarithmic term was consistently $-3/2$.\footnote{It was
later shown, however, that the logarithmic prefactor will  be equal
to $-1/2$ for  dilaton-gravity theories in four-dimensional spacetime
\cite{blah0}.}
This is a significant outcome, as it agrees precisely with the
analogous calculation made by Kaul and Majumdar
in a quantum-geometry context \cite{kaul}. Other support for 
this logarithmic correction and the  prefactor of $-3/2$ has
since followed 
\cite{blah1,blah2,blah3,blah4}.\footnote{Although
the prefactor has varied, the logarithm of the area
has turned up in other quantum-corrected treatments
of the black hole entropy. See, for example, Refs.\cite{dim,manns,med}.} 
\par
The focus of our  current study is to further
 test the applicability (from a black hole-duality perspective)
of   Carlip's  quantum-corrected    Cardy formula. 
For this purpose, we will be considering
 a  generic theory of 1+1-dimensional gravity
with coupling to a dilaton (i.e., auxiliary) field.\footnote{In
 spite of the claim of generality, it
will be  implied that the theory admits black hole
solutions, as well as a few other restrictions along the 
way.} 
The motivation for studying such a theory, in the context
of statistical entropy calculations,
 has been detailed
in the above discussion.
For some additional
background  on the  various aspects of  two-dimensional dilaton-gravity
theories, see Ref.\cite{germ} (and the citations within).
\par
The rest of this paper is organized as follows.
In Section 2, we begin by introducing the 1+1-dimensional model
of interest; after which, we discuss the   associated solution 
and thermodynamics
 at a classical level. This is followed by a 
{\it thermodynamic} evaluation  of the
leading-order quantum correction to the entropy. For this calculation,
we apply a formula that has recently
been derived by Das {\it et al}.  and  follows from
purely thermodynamic principles \cite{das}.
\par
In Section 3,  the quantum-corrected form of the Cardy formula 
(as derived by Carlip \cite{carlip}) is utilized for
an entropic calculation that is based strictly  on statistical-mechanical
principles.  To obtain the correct form of the Virasoro central charge,
as well as the eigenvalue of the zero-mode  generator, 
we apply a  methodology that was formalized 
by  Solodukhin \cite{sol}.  This author  demonstrated
that, for many (if not all) black holes, the  near-horizon geometry
 can  effectively be described
by  a  two-dimensional conformal field theory.
By way of analogy with this study, we are able to deduce the relevant Virasoro
parameters and then apply these to  Carlip's revised Cardy
formula.  The resultant form of the quantum-corrected
Cardy entropy is compared with the thermodynamic 
calculation of the prior section. Interestingly,  we find an apparent
discrepancy arising at the first perturbative order. 
\par
Finally,
Section 4 contains   a  summary and some closing  discussion.

\section{Quantum-Corrected Thermodynamic Entropy}

Since our current interest is in a  generic theory  of
1+1-dimensional (dilaton) gravity, let
us begin  by introducing an appropriate action:
\be
I={1\over 2G}\int d^2x \sqrt{-g}\left[D(\phi)R(g)+{1\over 2}
g^{\mu\nu}\nabla_{\mu}\phi\nabla_{\nu}\phi +{1\over l^2} V(\phi)\right].
\label{1}
\ee
Here, $G$ is a dimensionless measure of gravitational coupling
(i.e., the two-dimensional ``Newton constant''), $l$ is a 
fundamental constant of dimension length (for example, Planck's length),
while $D(\phi)$ and $V(\phi)$ are well-behaved but otherwise
 arbitrary functions of the
dilaton field. 
This is essentially the most general (diffeomorphism-invariant) action that 
contains at most
second derivatives of the  relevant fields: the metric and dilaton.\footnote{In
principle, the above action can effectively describe the  same gravity theory
 coupled  to an Abelian gauge field.  
For a gauge-invariant action, the Abelian 
sector can always
  be solved exactly in terms of only the dilaton and a conserved charge 
\cite{abelian1,abelian2}.
 Thus, the total action can consistently be re-expressed in the form
of Eq.(\ref{1}).} 
Note that an auxiliary field or  dilaton 
is a necessary element,  as the Einstein  tensor
identically vanishes in two dimensions of spacetime. 
\par
Although we are considering gravity from a two-dimensional
perspective, it is interesting to note that the above action
can often have  physical significance with regard to higher-dimensional
theories. For example, $D=\phi^2/4$ and $V=1$ corresponds
to the action obtained from the spherically symmetric
reduction of  3+1-dimensional Einstein gravity \cite{spher}.
Furthermore, such two-dimensional theories commonly
arise in  the near-horizon formulation of  
 near-extremal scenarios \cite{ext} and often have relevance
to string-theoretical models
\cite{youmy}.
\par
It is convenient to re-express the generic action in a form
for which the kinetic term is eliminated. This requires
the following reparametrization \cite{kun}: 
\be
\p=D(\phi),
\label{3}
\ee
\be
\g_{\mu\nu}=\Omega^2(\phi)g_{\mu\nu},
\label{2}
\ee
\be
\Omega^{2}(\phi)=\exp\left[{1\over 2}\int^{\phi}d\phi\left({dD\over d\phi}
\right)^{-1}\right],
\label{4}
\ee
\be
\V(\p)={V(\phi)\over\Omega^{2}(\phi)}.
\label{5}
\ee
It should be noted that our reparametrization
requires  $D(\phi)$ and its derivative to be non-vanishing
throughout the relevant manifold. 
\par
With the above field redefinitions, the action (\ref{1}) takes
on the following compact form:
\be
I={1\over 2G}\int d^2x \sqrt{-\g}\left[\p R(\g)+
{1\over l^2} \V(\p)\right].
\label{6}
\ee
\par
Given this apparent  simplification,  
it is not difficult to  obtain  the  general solution to
the reparametrized field equations. Moreover,
for the submanifold $x\geq0$,
 this
 solution can  readily be
expressed in a static, ``Schwarzschild-like'' gauge \cite{kuny}:
\be
\p={x\over l} \quad \geq 0, 
\label{7}
\ee
\be
ds^2=-\left(J(x)-2lGM\right)dt^2+ \left(J(x)-2lGM\right)^{-1}dx^2,
\label{8}
\ee
\be
J(x)=\int^{x/l}d\p\V(\p),
\label{9}
\ee
where $M$ (which is assumed to be non-negative)  
is a  constant of integration that can be identified
with the conserved mass of a black hole solution (assuming one exists).
\par
  In the subsequent
analysis, we will assume that the theory admits
black hole solutions for which the outermost horizon, $\p_o=x_o/l$,
is always  non-degenerate.
Generally speaking, one can locate an apparent
 black hole horizon by identifying
a hypersurface of vanishing Killing vector \cite{wald}. In
our case, this condition translates to the following
relation  \cite{mann,kuny}:
\be
J(x_o)-2lGM=0.
\label{10}
\ee 
\par
Next, let us consider the  black hole thermodynamic properties at the classical
level.
We can calculate the
 Hawking temperature ($T_o$) by analytically continuing to Euclidean spacetime
and then enforcing imaginary-time periodicity \cite{gh}. This process yields:
\be
T_{o}={1\over 4\pi l}\left. dJ\over d\p\right|_{\p_o}.
\label{11}
\ee
\par
In a 1+1-dimensional spacetime, there is no obvious definition for the
horizon area of a black hole.\footnote{Generally speaking, in a 
$p$+1-dimensional spacetime, a surface area can be regarded
as a ($p$-1)-dimensional measure of spatial extent. Alas,
this interpretation is ambiguous in the case of $p=1$.}
Hence,
the Bekenstein-Hawking area  law \cite{bek,haw} can not be exploited
in a straightforward manner.
However, we can still
evaluate the classical thermodynamic entropy ($S_o$) by 
considering the first law
of thermodynamics: $dM=T_{o}dS_{o}$.  Directly
applying Eqs.(\ref{10},\ref{11}) and then integrating,
we find: 
\be
S_{o}={2\pi\over G}\p_o,
\label{12}
\ee
where the integration constant has been set to zero
in accordance with the usual convention. It is interesting to note that:
\be
A_o= 4G S_{o} =8\pi\p_o
\label{area}
\ee
 can
now be interpreted as the effective ``area'' of the black hole
horizon. (Here, we have just applied the Bekenstein-Hawking entropic 
definition.)
\par
Next, we will proceed to evaluate
the leading-order quantum  correction to this classical  entropy.
On the basis of general thermodynamic arguments,
Das {\it et al}.  deduced that the  black hole entropy ($S$)
can be expressed via the following expansion  \cite{das}: 
\be
S= S_{o}-{1\over 2}\ln(C_oT_{o}^2)+...
\label{13}
\ee
where   ``...'' represents higher-order terms (with
regard to thermal displacements from equilibrium) and
$C_{o}$ is a dimensionless specific heat.\footnote{In order
to restore the proper dimensionality in Eq.(\ref{13}), 
 the quantity in the logarithm should be divided
by the square of Boltzmann's constant (which we have set equal to
unity, throughout).}  More
specifically:
\be
C_{o}= {\partial M\over \partial T_{o}}.
\label{14}
\ee
\par
Note that the above expression (\ref{13}), 
although quite general, has a limited
range of  validity \cite{das}.
In particular,  the equilibrium temperature ($T_o$)  must be significantly
larger than  the inverse of the natural length scale (in our case, $l^{-1}$).
 This condition rules out extremal and  near-extremal black
holes from further  consideration. Furthermore, Eq.(\ref{13}) can only 
be directly
applied if the specific heat is non-negative.  (Although this latter
constraint can typically  be circumvented  via a
suitable regulatory parameter  \cite{das}.)
 \par
Up until  now, we have been  treating the
dilaton potential ($V(\phi)$ or $\V(\p)$) as generically as possible. 
However, for illustrative
purposes, it is instructive if this potential is given a more
specific form. Let us thus consider:
\be
\V(\p)= \gamma\p^a ,
\label{15}
\ee
where $a$ and $\gamma$ are dimensionless, non-negative, model-dependent
 parameters.
Notably, this ``power-law'' potential can  correspond to 
a Weyl-rescaled CGHS model (for $a=0$ and $\gamma=1$) \cite{cal, cad2}, or
a dimensionally reduced
 BTZ black hole (for $a=1$ and $\gamma=2$) \cite{btz,ao}.
More generally, such a model is  (after an appropriate rescaling)
capable of describing the 
  near-horizon geometry of a single-charged
 dilatonic black hole, a multi-charged stringy black hole, or
a dilatonic $p$-brane \cite{youmy}.\footnote{Note that, 
by restricting $a\geq0$,
 we have eliminated dimensionally reduced, 
spherically symmetric  Einstein  gravity from present considerations. 
After a suitable  reduction and  field redefinition, $d$-dimensional Einstein
gravity is  described
by a power-law potential with $a=-1/(d-2)$ and $\gamma=(d-3)/(d-2)$
\cite{kun2}. Although such theories are interesting,
 this restriction is necessary  to avoid the complication
of a negative specific heat.}
 \par
Given this power-law form for the potential,
we can apply Eqs.(\ref{10},\ref{11},\ref{14}) to obtain:
\be
M={1\over 2lG}{\gamma\over a+1} \p_o^{a+1},
\label{16}
\ee
\be
T_{o}={\gamma\over 4\pi l}\p_o^a,
\label{17}
\ee
\be
C_{o}={2\pi\over G a}\p_o.
\label{18}
\ee
Note that the specific heat is always positive; cf. Eq.(\ref{7}).
\par
Substituting the above results into Eq.(\ref{13}),
we have (up to constant terms and higher-order corrections)
the following outcome:
\be
S=S_o-{2a+1\over 2}\ln(S_o)+....
\label{19}
\ee
Such a  logarithmic correction to the ``area law''  is a  familiar 
occurrence. See, for instance, the calculations
of Kaul  and Majumdar in a quantum-geometry context
\cite{kaul}. (Also see 
Refs.\cite{carlip,blah0,blah1,blah2,blah3,blah4,dim,manns,med,das}.)

\section{Quantum-Corrected Cardy Entropy}

In the current section, we reconsider the black hole entropy with respect to
a generic 1+1-dimensional theory. 
This time, however, the calculation will be based  on the
principles of
 statistical mechanics.
 In particular, we will apply  the Cardy formula \cite{car};
including Carlip's leading-order  quantum correction \cite{carlip}.
\par
The Cardy formula follows from a saddle-point approximation
of the  partition function  for a two-dimensional conformal
field theory. This leads to the theory's density of states (i.e., the
exponential of the entropy), which is related to
the partition function  by way of a Fourier transform.
\par
For further details on the derivation and  significance of
 the Cardy formula, see  
 Ref.\cite{carlip2}.  Here, we will simply quote the result
of Carlip's  quantum-corrected version \cite{carlip}:
\be
\rho(\Delta)\approx \left({c\over 96 \Delta^3}\right)^{1/4}
\exp\left[2\pi \sqrt{{c\Delta\over 6}}\right],
\label{20}
\ee
where $c$ is the ``central charge'' of the conformal theory,
$\Delta$ is the eigenvalue of the zero-mode Virasoro generator
(noting that a  conformal field theory realizes a
representation of the quantum Virasoro algebra \cite{cft}) and $\rho(\Delta)$
is the corresponding density of states. Note the approximation sign,
indicating that only the classical and leading-order quantum contributions  
have been considered.
\par
If we are to apply this formula to generic  1+1-dimensional gravity,
it is necessary to   
 show that the model under consideration
is dual to a two-dimensional conformal field theory. That is,  show
that the symmetry generators ($L_n$) of an effective gravitational action 
are capable of satisfying the standard Virasoro algebra \cite{cft}:
\be
\left[L_n,L_m\right]=(n-m)L_{n+m}+{c\over 12}n(n^2-1)\delta_{n+m,0}.
\label{21}
\ee
Even after accomplishing  this non-trivial  task, it would
still be necessary to obtain explicit forms of the key ingredients;
namely, $c$ and $\Delta$. (The latter being the eigenvalue of $L_{0}$.)
\par
Fortunately,  both of these formidable steps have essentially been done
for us in a prior work  by Solodukhin \cite{sol}. Next, let us give
a brief account of this treatment.
\par  
Solodukhin began in Ref.\cite{sol} by showing  that many (if not all)
gravity theories describing a black hole will provide a
realization of the Virasoro algebra in a region  sufficiently
close to the horizon.  The author went on to examine
 four-dimensional Einstein gravity\footnote{Subsequently in the
same paper \cite{sol},
 this analysis was generalized to 
 $d$-dimensional Einstein theory, such that $d\geq 3$. It is,
however, necessary to add a (negative) cosmological constant term in
the $d=3$ case, in order to obtain a non-trivial
solution of the field equations.}
  on a class of 
spherically symmetric metrics. After a suitable process of dimensional
reduction and field redefinition, it was then demonstrated that
the effective action takes on a Liouville-like form \cite{loo}.
Significantly, Liouville theory is indeed a two-dimensional
 conformal field theory and
often  plays a significant role 
in describing  black hole geometries near a boundary (particularly
in 2+1-dimensional gravity)  \cite{cous}.
Moreover, the  
 resultant effective action only differed from
 standard Liouville theory   in the
precise form of
its  dilaton potential, which
 turned out  to be  inconsequential in a near-horizon  regime.
(Rather, this potential is suppressed by the horizon red-shift
factor).
\par
Ultimately, Solodukhin exploited this duality to
obtain the appropriate values of the relevant Virasoro parameters, $c$ and 
$\Delta$.\footnote{Whereas the central charge ($c$) followed directly
with the identification of the Virasoro algebra, this was not the case
for   the  zero-mode eigenvalue 
($\Delta$). This  latter consideration   was  complicated  by virtue of a 
vanishing $L_{0}$ for
a strictly classical configuration. Solodukhin remedied this situation
by assuming periodicity  and then  imposing 
suitably chosen  boundary 
conditions on the near-horizon dilaton field \cite{sol}.}
With these identifications, 
the Cardy formula could directly be  applied to evaluate
the  entropy of the  originating theory (i.e., 3+1-dimensional Einstein
gravity). Remarkably, the standard Bekenstein-Hawking result was
exactly reproduced.  (It was also reproduced  for  similar treatments of
higher-dimensional Einstein gravity
and  2+1-dimensional anti-de Sitter gravity \cite{sol}.)
\par
We can directly apply the results of the Solodukhin treatment, 
provided that our 1+1-dimensional gravity model, as described by
Eq.(\ref{1}) or Eq.(\ref{6}), can be re-expressed  
in terms of a Liouville-like model. As it turns out, this can be accomplished 
with the following field redefinitions:
\be
\pp={2\over Gq}\p,
\label{22}
\ee
\be
\G_{\mu\nu}=\exp\left[-{2\over q}\pp\right]\g_{\mu\nu},
\label{23}
\ee
where $q$ is an arbitrary, dimensionless parameter.
\par 
With this additional reparametrization, the action (\ref{6})
adopts  the desired (Liouville-like)  form:
\be
I=\int d^2x \sqrt{-\G}\left[{q\over 4}\pp R(\G)+ {1\over 2}\G^{\mu\nu}
\nabla_{\mu}\pp\nabla_{\nu}\pp +\VV(\pp)\right],
\label{24}
\ee
where the revised dilaton potential is given by:
\be
\VV(\pp)={1\over 2G}e^{{2\over q}\pp}\V(\p(\pp)).
\label{25}
\ee
As mentioned  above, the  near-horizon geometry (which determines
the Virasoro algebra) is insensitive to
the precise form of the reparametrized dilaton potential  \cite{sol}.
It is only necessary that this potential
remains non-singular (at least near the horizon),
which is trivially the case.
\par
In direct analogy to the Solodukhin program, we are able to deduce
that the near-horizon form of Eq.(\ref{24}) (and, hence, the near-horizon
form of
generic 1+1-dimensional gravity) can be described by
a conformal field theory. Moreover, the relevant Virasoro parameters
are given as follows \cite{sol}: 
\be
c=3\pi q^2,
\label{26}
\ee
\be
\Delta={\pp_o^2\over 2\pi}= {2\over \pi G^2 q^2}\p_o^2,
\label{27}
\ee
where $\pp_o$ is (of course) the horizon value of $\pp$.
\par 
Substituting the above results into Eq.(\ref{20}) for
the density of states, we have:
\be
\rho \approx \sqrt{{3\over 2}}c\left({G\over 6\p_o}\right)^{3/2}
\exp\left[{2\pi\p_{o}\over G}\right].
\label{28} 
\ee
\par
Next, we make the usual identification in defining the entropy
($S=\ln\rho$) and also apply the two-dimensional
 analogue of the Bekenstein-Hawking
law:  $S_{o}=A_{o}/4G=2\pi\p_o/G$  (cf. Eq.(\ref{area})).
This leads to the following expansion:
\be
S=S_{o}-{3\over 2}\ln(S_o)+\ln(c)+...,
\label{29}
\ee
where ``...'' represents both higher-order corrections and constant
terms. Clearly, the classical thermodynamic result has been reproduced
at the lowest order. 
Although an anticipated outcome, this 
had not  yet been explicitly
verified for a generic two-dimensional theory. 
(Note that the arbitrary parameter, $q$, has been
effectively canceled off; at least at the classical level.) 
\par
 Let us now assume that the central charge is ``universal'' in the
sense that $c$ is
 independent of $\p_o$.\footnote{This assumption follows
from the usual notion that the central charge is a measure
of the number of massless particle species \cite{car}.}  
In this case,  we have also substantiated
Carlip's claim of a leading-order correction that is, up to an irrelevant
 constant,
 just  the
logarithm of the ``area'' \cite{carlip}. 
  Furthermore,
we have also obtained the anticipated prefactor of
$-3/2$.
\par
In spite of the success of this treatment, we unfortunately
observe a discrepancy between this state-counting 
calculation and the  thermodynamic calculation of
Section 2 (cf. Eq.(\ref{19})).  From a purely thermodynamic 
perspective, the leading-order
logarithmic correction has a prefactor that is, in general,
{\bf not} equal to $-3/2$. For a power-law potential in particular
(cf. Eq.(\ref{15})), we found that the prefactor only equals
this desired value for the special case of $a=1$.
This choice of $a$ describes a theory of Jackiw-Teitelboim gravity
\cite{jt} or, from a higher-dimensional perspective,
the dimensionally reduced BTZ black hole \cite{btz,ao}.
We will endeavor to rationalize this inconsistency in the concluding section.

\section{Conclusion}

In summary, we have considered a very general theory of 1+1-dimensional
gravity with coupling to an auxiliary (dilaton) field.
We began by demonstrating a  procedure of field redefinition
that conveniently eliminates the kinetic term from
the generic action \cite{kun}.  After which, the classical solution
was  presented along with  the associated thermodynamic
properties of the (assumed) black hole  horizon.
Applying a calculation by Das {\it et al}. \cite{das}, we
were then able to deduce the leading-order correction
to  the classically defined  entropy.  We found that this correction is
(up to a constant) just the logarithm of the ``area'' (i.e.,
the two-dimensional analogue of area as based on the Bekenstein-Hawking
 law \cite{bek,haw}).
 This outcome is in agreement with prior
calculations of the quantum-corrected black hole
entropy (for instance, Ref.\cite{kaul}). 
\par
Following this purely thermodynamic treatment, we 
proceeded to consider the black hole entropy from a statistical-mechanical
perspective.  For this purpose, we utilized a methodology
that has been developed by Solodukhin \cite{sol}. 
The premise of this program is
that  gravity theories admitting
a black hole solution will (typically) 
have a near-horizon duality with a two-dimensional
conformal theory. On the basis of this correspondence, it
is possible to identify the Virasoro parameters that are
needed in the Cardy  formulation of  the density of states \cite{car}.
By analogy with Ref.\cite{sol}, we were able to identify these parameters
and subsequently calculate the statistical entropy.  At the lowest order,
this  gave us back the  classical thermodynamic result; thus justifying
the implied choice of  boundary conditions (in evaluating the eigenvalue
of the zero-mode Virasoro operator \cite{sol}).
\par
Along with the classical consideration,
 we  applied  these Virasoro parameters 
in calculating     the leading-order correction to
the  Cardy entropy; the generic form of this correction having recently been
 derived by Carlip \cite{carlip}.  As in the purely thermodynamic calculation,
we found the correction to be given by the  logarithm of the ``area''.
Moreover, in this state-counting calculation, we  demonstrated
that the  logarithmic prefactor takes on a universal value
of $-3/2$. Although this particular value is
supported by the literature \cite{kaul,carlip}, it conflicts
with the outcome of our proceeding analysis.
From a thermodynamic viewpoint, this prefactor is decidedly
model dependent and only takes on a value of $-3/2$
for a limited range  of two-dimensional actions; for instance,
that which  describes Jackiw-Teitelboim theory \cite{jt}.
\par
In an attempt to rationalize this bothersome discrepancy,
we turn to a pair of conspicuous  limitations in the  formal
treatment of Section 3. First of all,  we used a central
charge that is purely classical in  its origins. 
In general, this central charge should include a quantum correction,
which would change the exponent in the Cardy formula (\ref{28})
from its classical value of $2\pi\p_o/G$. However, it
has been convincingly  argued by Carlip (see Appendix B of Ref.\cite{carlip})
that the first observable quantum effects  of
a corrected central charge 
 will come
at the order of inverse area.  That is,  at the order of $\p_o^{-1}$ in
generic 1+1-dimensional gravity.\footnote{More support along this line
 has since followed \cite{marco}.
In the cited study, it was demonstrated that,
 at least for 1+1-dimensional theories that are asymptotically
Jackiw-Teitelboim,
 quantum-gravity effects will show up  at the order
 of  $\p_o^{-2}$. This analysis was based on a perturbative expansion of 
a suitable target-space metric.}  
  Such corrections should, therefore, have no 
repercussions on the prefactor of the logarithmic term.
\par
Secondly, we again point out that  our application of the Cardy formula
was based on a near-horizon duality between the original
gravity theory and a conformal theory field.
Thus,  the Cardy formula was only  capable of counting degrees
of freedom that live at (or very close) to the  black hole horizon.
On this basis, the observed discrepancy implies
that other degrees of freedom may  become important as the
quantum aspects of the theory are more closely probed.
Interestingly, this viewpoint coincides with Smolin's
notion of both a  weak and  strong version of the
 holographic principle \cite{smolin}.
That is, the number of degrees of freedom  in a black hole's interior
 is not necessarily in agreement with the degrees
of freedom that are accessible to an external observer. 
\par
In conclusion, the microscopic origin of black hole entropy
remains an intriguingly open question, which will undoubtedly 
be the subject of many more future investigations.

\section{Acknowledgments}

The author would like to thank V.P. Frolov for
helpful conversations.

\end{document}